\documentclass[12pt,preprint]{aastex}

\slugcomment{Accepted by ApJL, May 2011}

\shorttitle{Supra-Arcade Downflow Distributions}
\shortauthors{McKenzie \& Savage}

\begin{document}


\title{Distribution Functions of Sizes and Fluxes
  Determined from Supra-Arcade Downflows }


\author{D. E. McKenzie}
\affil{Department of Physics, PO Box 173840, Montana State University,
    Bozeman, MT 59717-3840}

\and

\author{S. L. Savage}
\affil{NASA/Goddard Space Flight Center (Oak Ridge Associated Universities),
	8800 Greenbelt RD, Code 671, Greenbelt, MD 20771}

\begin{abstract} 
The frequency distributions of sizes and fluxes of supra-arcade 
downflows (SADs) provide information about the process of their 
creation.  For example, a fractal creation process may be expected to 
yield a power-law distribution of sizes and/or fluxes.  We examine 120 
cross-sectional areas and magnetic flux estimates found by 
\citet{SavageMcKenzie_11} for SADs, and find that (1) the areas are 
consistent with a log-normal distribution, and (2) the fluxes are 
consistent with both a log-normal and an exponential distribution. 
Neither set of measurements is compatible with a power-law 
distribution, nor a normal distribution. As a demonstration of the 
applicabiilty of these findings to improved understanding of 
reconnection, we consider a simple SAD growth scenario with minimal 
assumptions, capable of producing a log-normal distribution.
\end{abstract}


\section{Introduction}

Supra-arcade downflows (SADs) are downward-moving features 
observed in the hot, low-density region above posteruption 
flare arcades.  Initially detected with the {\em Yohkoh} 
Soft X-ray Telescope (SXT) as X-ray-dark, blob-shaped 
features, downflows have since been identified in many 
flares, with a variety of instruments 
\citep{McK_00,Innes_EA03a,Khan_EA07,SavageMcKenzie_11}. 
Descriptions of their characteristics, with example 
movies, are provided by \citet{McK_00}; discussion of 
semi-automated routines for detecting and tracking SADs 
can be found in \citet{McKenzieSavage09}. The darkness of 
these features in X-ray and extreme-ultraviolet (EUV) 
images and spectra, and their lack of absorption 
signatures in EUV, indicate that the SADs are best 
explained as pockets of very low plasma density, or {\em 
plasma voids} \citep[see especially][]{Innes_EA03a}. 
Although SADs of the plasma void type were the first to be 
identified \citep{McK_HSH99}, numerous instances of {\it 
shrinking loops} have also been reported 
\citep{McK_00,MR2K}.  \citet{Savage_EA10} introduced the 
moniker SADLs (supra-arcade downflowing loops) for 
downflows of this latter type. Interpreting both types as 
shrinking post-reconnection flux tubes, 
\citet{McKenzieSavage09} utilized the measured 
cross-sectional area of plasma voids and an estimate of 
the magnetic flux density to assign an estimated magnetic 
flux to each shrinking tube.  The flux estimated in this 
way, $\sim 10^{18}$ Mx, is notably similar to the flux 
reported by \citet{Longcope_EA05}, found by a different 
means.

In \citet{SavageMcKenzie_11} the authors presented 
measurements of SADs and SADLs sampled from a large number 
of solar flares observed with {\em Yohkoh}/SXT, TRACE, 
{\em Hinode}/XRT, and SoHO/LASCO.  In the present Letter, 
we attempt to infer from the frequency distributions of 
SAD sizes and magnetic fluxes information about the 
process(es) responsible for creation of the SADs. Such a 
study is the natural extension of the ensemble 
characteristics presented by \citet{SavageMcKenzie_11}, 
and the findings have implications for an improved 
understanding of the formation of SADs, and perhaps for 
numerical modeling of patchy magnetic reconnection. For 
example, the fractal current sheet proposal of 
\citet{ShibataTanuma_01} would tend to yield copious 
numbers of smaller and smaller magnetic islands forming in 
the current sheet in a scale-invariant cascade.  The 
frequency distribution of sizes, and perhaps fluxes, of 
magnetic islands in such a fractal current sheet is 
expected to tend toward a power-law distribution.  
\citet{Nishizuka_EA09} analyzed small-scale brightenings 
in the 14 July 2000 `Bastille Day' flare, and found that a 
power-law distribution produced a good fit to the ``peak 
intensity, duration, and time interval'' of the time 
profiles of the brightenings.  \citet{Nishizuka_EA09} 
interpreted the power-law distribution as possible 
evidence supporting the fractal current sheet concept.  
Similarly, \citet{Aschwanden_EA98} fitted curves to a very 
large sample (tens of thousands) of hard X-ray pulses, 
radio Type III bursts, and decimetric pulsations and 
spikes, all interpreted as signatures of elementary energy 
release events.  \citet{Aschwanden_EA98} deduced for each 
burst the peak energy dissipation rate and then explored 
the frequency distribution of those energy fluxes.  Those 
authors found that a power-law distribution provided a 
satisfactory representation of the whole sample; on the 
other hand, analysis of bursts in individual flares 
suggested that while power-law distributions provided good 
fits for the hard X-ray pulses and Type III bursts in many 
flares, the decimetric pulsations and decimetric 
millisecond spikes were better fit by exponential 
distribution functions.  \citet{Aschwanden_EA98} suggested 
that the variations in frequency distribution may be 
related to differences in the relevant emission sources.

In a model of island formation in current sheets, 
\citet{Fermo_EA10} determined an analytic expression for 
the expected frequency distribution of fluxes and radii of 
islands, including consideration of possible merging of 
the islands.  For the case of very little merging, 
\citet{Fermo_EA10} found that the frequency distribution 
resembles a decaying exponential.  In a separate study, 
\citet{Drake_EA10} simulated islands in stacked current 
sheets in the heliosheath, a simulation which included 
merging of the islands into `bubbles'.  
\citet{Opher_EA11} presents a description of the evolution 
of these `bubbles', and finds (non-rigorously) that the 
frequency distribution of magnetic field intensities 
within the bubbles roughly resembles a log-normal 
distribution.

In the following sections we compare the measured areas 
and fluxes from \citet{SavageMcKenzie_11} to some common 
statistical distribution functions in an effort to draw 
inferences about their formation. The final distribution 
of SAD sizes is of course affected by the processes of 
creation and growth, and may also be affected by 
subsequent fragmentation into smaller voids, and merging 
of voids. Although fragmentation and merging may be 
possible for SADs, these events have not been unambiguosly 
observed in coronal images to date. Future observations 
may provide evidence for fragmentation and/or merging, 
allowing empirical estimation of their relative 
importance; we will not consider these effects herein. 
Pursuant to the interpretation of the SADs/SADLs as 
reconnection products, it is thus conjectured that the 
sizes and magnetic fluxes yield information about the rate 
of reconnection in the individual flux transfer events.  
We find that a log-normal distribution appears to best fit 
the SAD sizes, and is also consistent with the SAD fluxes. 
We provide a simple interpretation, with minimal 
assumptions, to indicate one way a log-normal distribution 
might result from the growth of the plasma voids.

\section{Measurements and Distributions}

The measurements of \citet{SavageMcKenzie_11} include SADs 
and SADLs from more than one instrument, and more than one 
flare.  In that work, the SADLs in any one flare were all 
assigned the same diameter, a choice that was necessitated 
by the data.  In the present work, we consider only the 
SADs (i.e., downflows of the {\it plasma void} type), 
since the areas were determined from signal thresholds in 
the images.  Furthermore, we examine only the 120 SADs 
found with {\em Yohkoh}/SXT, in order to avoid 
instrument-dependent variations in the distributions that 
arise when the data from multiple telescopes are combined.
The angular resolution for most of the SXT observations is 
4.91 arcseconds per pixel (SXT's ``half resolution''), 
corresponding to a area unit of approximately 12.5 Mm$^2$ 
per pixel.  SADs in the 26 June 2001 flare were analyzed 
with ``full resolution'', 2.455 arcsec/pixel.  For each of 
the SADs discussed herein, the area is found by averaging 
multiple measurements made during the SAD's lifetime.  

The TRACE observations listed by \citet{SavageMcKenzie_11} 
have the benefit of higher spatial resolution, but 
comprise a much smaller sample --- all 23 SADs are from a 
single flare. On the other hand, the SXT data have the 
benefit of a larger sample, but are drawn from 16 flares 
(listed in Table 1).  In order to produce a sample large 
enough for a meaningful examination of the frequency 
distribution, we combine the 120 SXT SADs into a single 
set, even though this introduces two related assumptions.  
First is the assumption that the spectrum of SAD sizes 
and/or fluxes is the same for all the flares. To a degree 
this seems reasonable, as it supposes that the processes 
which add/remove flux to a given flux tube are the same in 
each flare.  We presume that each flux tube is created by 
a burst of 3-D reconnection, and that each burst is 
independent of other bursts.  Thus the essence of the 
primary assumption is that reconnection proceeds in a 
similar way in each flare, or at least that the products 
of reconnection are similar in each flare.

The second assumption is that SADs are {\em detected} 
equally well in all the flares.  All SADs were initially 
found with an automated detection program, and had signals 
darker than their surroundings by at least 1.1 standard 
deviations.  Each was subsequently verified by visual 
inspection.  Owing to the contrast enhancement performed 
as part of the data preparation, there is no statistical 
correlation between detection threshold and measured SAD 
area.  As a test of the effect on detectability 
due to the brightness of background signal in the images, 
we asked if smaller SADs are detected less frequently in 
more intense flares. To examine this question we grouped 
the SADs into four bins, according to the GOES 
classification of the flares.  The bins are as follows: 
Group 1 includes SADs from flares with GOES C5.0 through 
M1.9; Group 2 includes GOES M2.4--M4.4; Group 3 includes a 
single flare with GOES M5.2; and Group 4 includes GOES 
M7.5--X1.2.  The 26 June 2001 SADs are not included in 
this binning, since that event occurred beyond the limb 
and did not have a recorded GOES classification. Examining 
the frequency distributions in these subsets, we see no 
clear evidence for a systematic observational bias --- the 
frequency distributions for the four groups are roughly 
similar, with reasonable variation due to the small 
samples.  Importantly, small SADs are found both in the 
largest and smallest flares.  Therefore we proceed with 
analysis of the combined set of SAD measurements, although 
the results should be interpreted with consideration of 
the underlying assumptions.

For the set of SAD area measurements, and for the set of 
SAD flux estimates, we show the frequency distributions in 
Figures 1(a) and 2(a).  As also mentioned in 
\citet{SavageMcKenzie_11}, the distributions show a 
greater number of smaller sizes/fluxes than larger ones.  
The downturn of frequency distribution for the smallest 
size/flux bins is characteristic of a log-normal 
distribution.  Plots of the frequency distributions for 
$ln($area$)$ and $ln($flux$)$ indicate normal-like 
distributions as well (and thus suggest log-normal 
distributions of size and flux), but are omitted here in 
the interest of space.  In Figures 1(a) and 2(a), a 
log-normal distribution is overplotted for comparison; 
Figure 2(a) also displays an exponential curve for 
comparison.  The shape of the log-normal curve is defined 
by the mean and standard deviation of the data, and the 
exponential curve is defined fully by the mean of the 
data. Therefore the curves in Figures 1(a) and 2(a) should 
be considered fits only to the extent that they produce 
the same mean, standard deviation, and total number of 
SADs as the data.  For a more quantitative diagnostic of 
the distributions we construct the cumulative distribution 
of the data in Figures 1(b) and 2(b), which are then 
compared to the analytic cumulative distribution functions 
(CDFs) of four common statistical distributions: normal, 
log-normal, exponential, and power-law. In each figure, 
the cumulative distribution of the data sample is shown as 
diamond-shaped symbols.

\begin{figure}[ht]
\plottwo{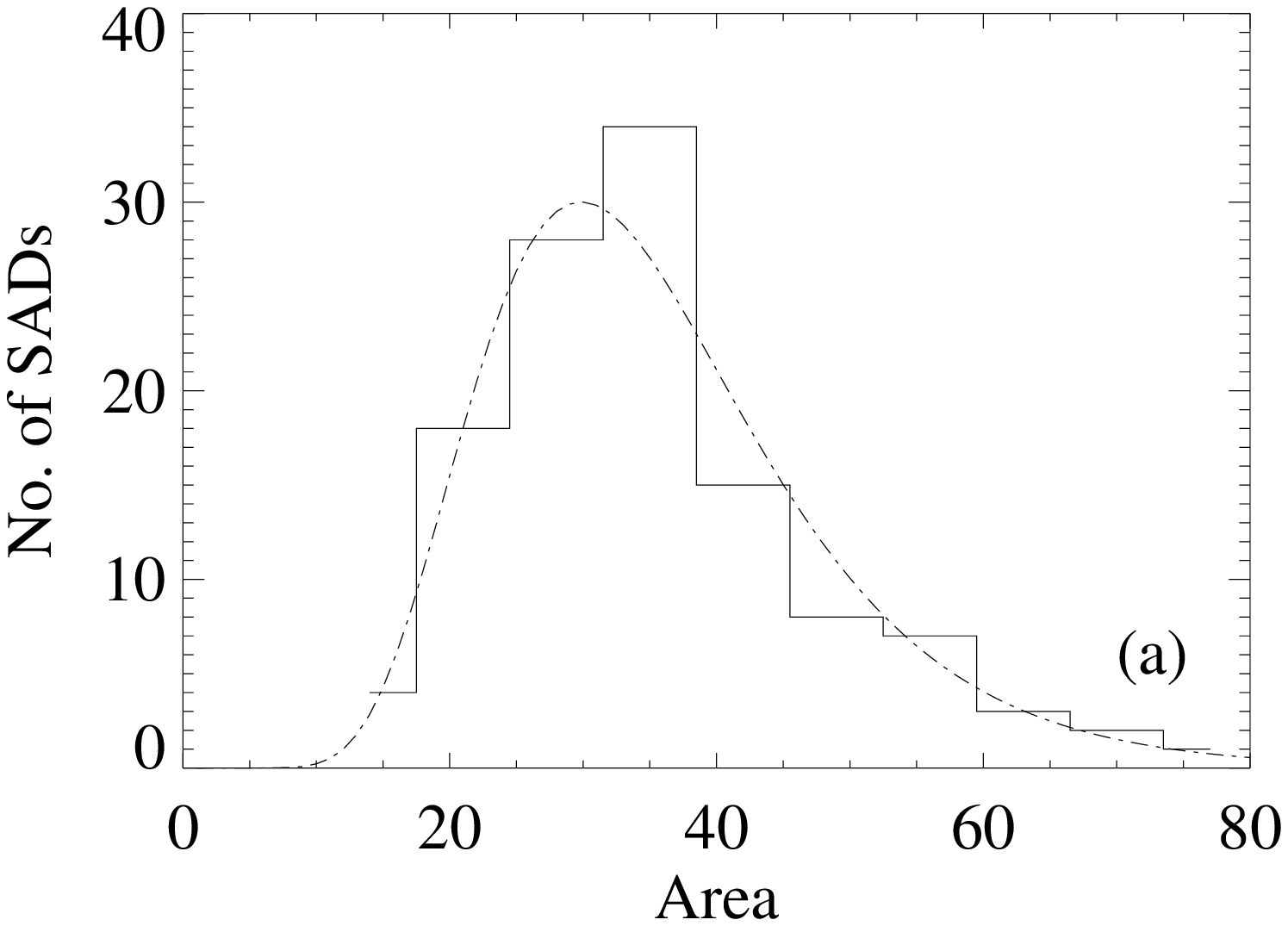}{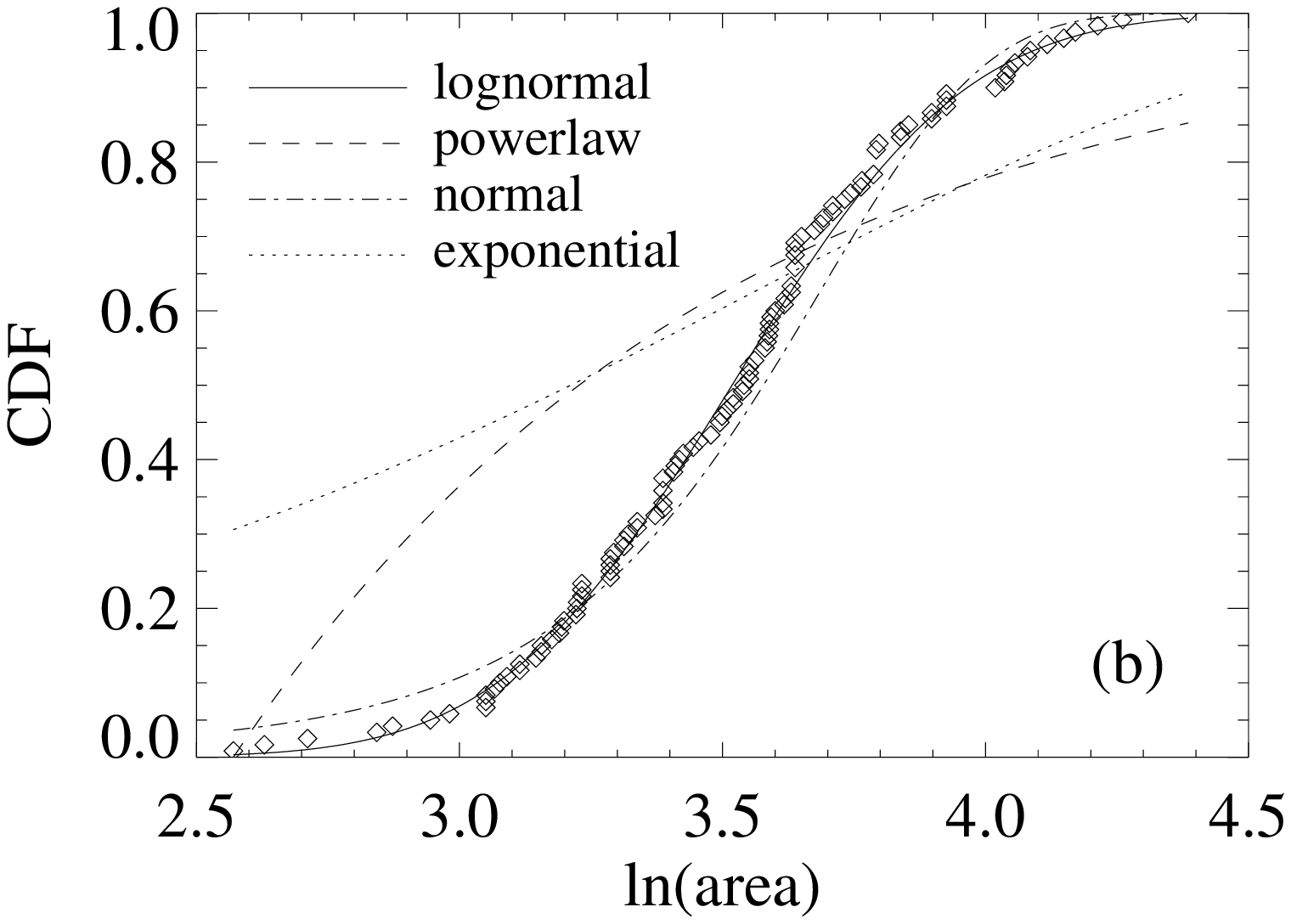}
\caption{Distribution of SAD areas.  The left-hand panel 
(a) shows the measured frequency distribution of SAD 
sizes, in units of Mm$^2$.  A log-normal distribution 
curve is overlaid for comparison.  The right-hand panel 
(b) displays the cumulative distribution of SAD sizes as 
diamond-shaped symbols, overplotted with theoretical 
CDFs.
\label{SXTareas}}
\end{figure}

\begin{figure}[ht]
\plottwo{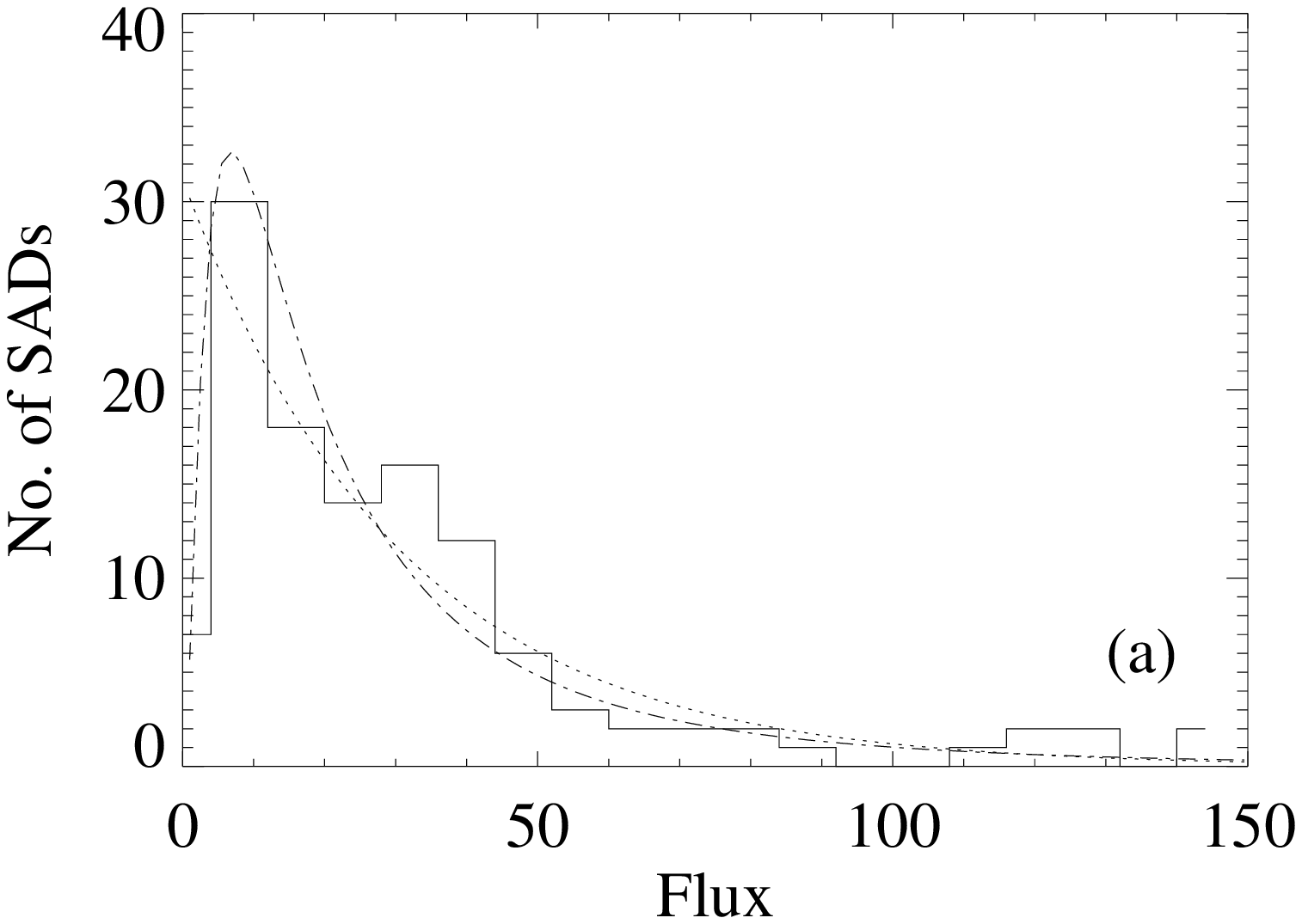}{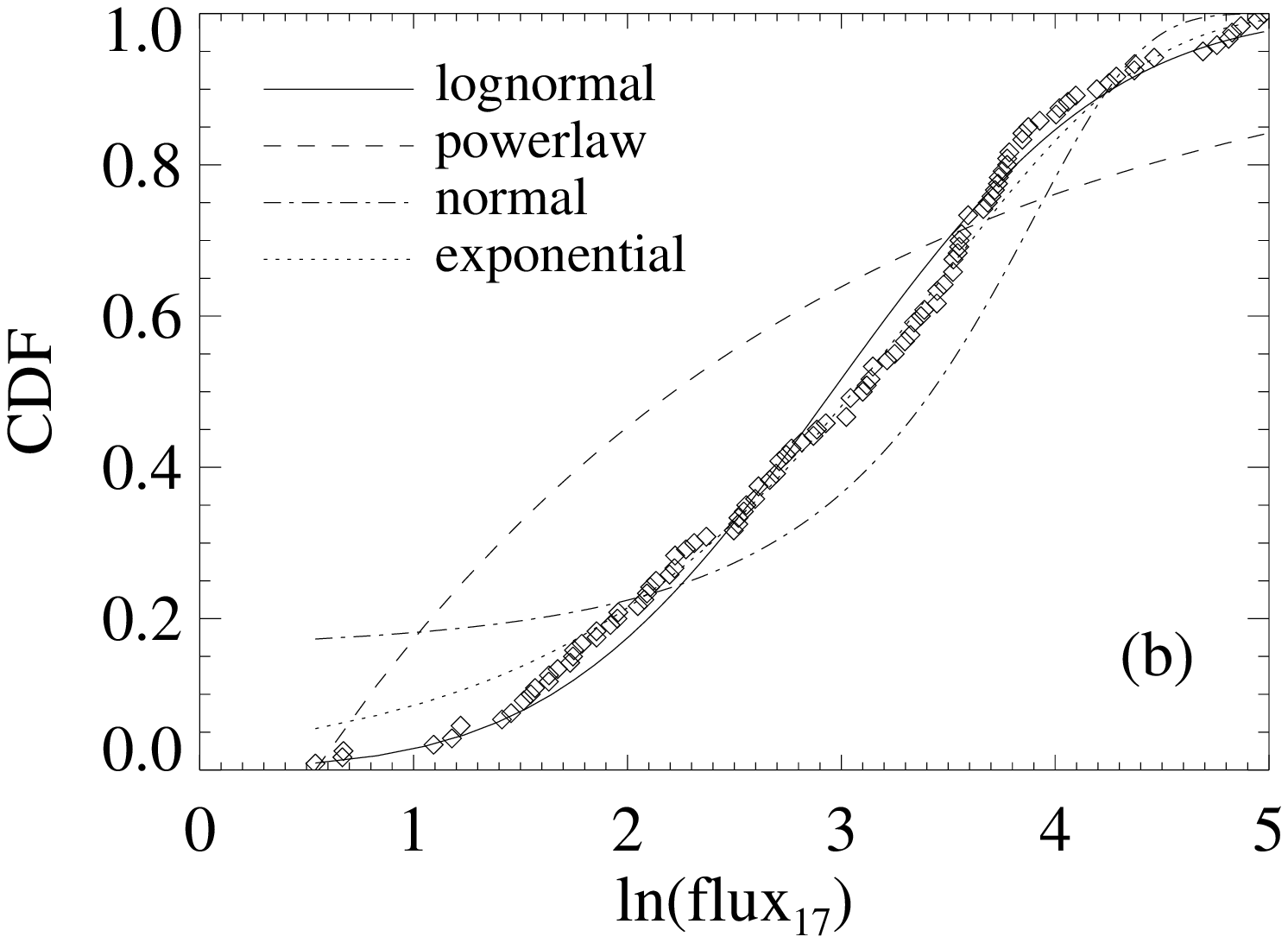}
\caption{Distribution of estimated SAD fluxes.  The 
left-hand panel (a) shows the frequency distribution of 
fluxes.  Overplotted curves show a log-normal 
distribution (dash-dot) and an exponential distribution 
(dotted) for comparison.  The right-hand panel (b) 
displays the cumulative distribution of fluxes as 
diamond-shaped symbols, overplotted with theoretical 
CDFs.  In both panels, the fluxes have been normalized by 
$10^{17}$ Mx.
\label{SXTfluxes}}
\end{figure}

To test the goodness of fit for each of the proposed CDFs, 
we employed the Kuiper variant of the Kolmogorov-Smirnov 
test.  According to \citet{Press_EA}, the traditional 
Kolmogorov-Smirnov tends to be most sensitive to 
deviations in the middle of a CDF's range, and least 
sensitive at the extrema.  We elected to use the Kuiper 
variant in order to preserve sensitivity to deviations 
throughout the full range of the CDF.  The Kuiper 
significances of the CDF comparisons are listed in Table 
2.  As in the traditional Kolomogorov-Smirnov test, small 
values of the significance indicate that the CDFs are 
incompatible.  From these significances, we infer the 
following: (1) The SAD areas are consistent only with a 
log-normal distribution.  (2) The SAD fluxes are 
inconsistent with a normal or power-law distribution, but 
consistent with both log-normal and exponential 
distributions.

As an additional test, we repeated the area analysis 
omitting the SADs measured in the 26 Jun 2001 flare.  
This flare contributes the majority of SAD areas smaller 
than 20 Mm$^2$, and therefore one may reasonably ask if 
the overall distribution is affected by their 
inclusion/exclusion.  We find that the result is 
unchanged: the areas of the other 111 SADs are consistent 
with a log-normal distribution (with Kuiper significance 
0.4), and the other distributions are ruled out completely 
(significances between $5\times10^{-4}$ and 
$6\times10^{-31}$).

Finally, we examined the 25 SADs from the 20 Jan 1999 
flare (i.e., Group 3).  Though the sample is much smaller, 
we find that the areas are consistent with a log-normal 
distribution (significance 0.3), and less likely to be 
normally distributed (signif. $8\times10^{-2}$). Power-law 
and exponential distributions are discounted with 
significances $2\times10^{-3}$ and $6\times10^{-8}$, 
respectively.

Although the TRACE data of \citet{SavageMcKenzie_11} 
provide size and flux estimates for some 23 SADs, the 
cumulative distribution curves of the TRACE data alone are 
very noisy, due to the sparseness of the sample.  The 
TRACE measurements indicate that SADs smaller than SXT's 
resolution exist, so one can conclude that the frequency 
distribution below the range shown in Figure 2 is 
non-zero. However, combination of the SXT and TRACE 
measurements is made difficult by the vast differences in 
angular resolution and dynamic range (i.e., contrast), 
which result in instrument-specific observational biases: 
as discussed in \citet{McKenzieSavage09}, the SAD areas 
measured from SXT may be systematically overestimated. In 
contrast, we find no reason to expect the flux density 
(i.e., magnetic field strength) to be limited by the 
observations. This is borne out by the observation that 
the 23 TRACE areas (ranging 1.9--12 Mm$^2$) and the 120 
SXT areas (13--80 Mm$^2$) do not appreciably overlap, 
whereas the flux ranges overlap completely ($[3.9-45] 
\times 10^{17}$ Mx for TRACE, $[1.7-145] \times 10^{17}$ 
Mx for SXT).  Thus, although the range of areas in the 
sample of 120 SADs is constrained by the SXT resolution, 
there appears to be no methodological limit on the SAD 
fluxes.

\section{Discussion}

We propose that the distribution of sizes and fluxes 
reveals clues about the process which forms the SADs.  
Pursuant to the interpretation of SADs as reconnection 
products, the distribution of sizes and fluxes indicate 
the nature of some parameter(s) of localized (patchy \& 
bursty) reconnection.  As an extreme example, consider a 
scenario in which all the SADs were found to have 
identical sizes and fluxes.  In such a case --- which is 
not consistent with the present measurements --- the 
implications for reconnection would be easy to define.  
In the present case interpretation is not as trivial, but 
is still possible.  In the case of a log-normal 
distribution, for instance, it is not the sizes of SADs 
that are distributed normally, but the logarithm of the 
sizes.  This can arise if, say, a reconnecting patch in 
the current sheet experiences growth at a rate that is 
proportional to the size of the patch:

\begin{equation}
\frac{dX}{dt} = kX~~.
\end{equation}
Thus the growth coefficient $k$ is given by
\begin{equation}
\frac{d~ln(X)}{dt} = k ,
\end{equation}
and one can assume that the size is allowed to grow for some
interval $\tau$ such that
\begin{equation}
X(\tau) = X_0 e^{k\tau}~~.
\end{equation}
This argument follows exactly the discussion of 
\citet{Koch_66}, and requires each patch to grow from some 
initial ``seed" size $X_0$.  We find that $X(\tau)$ will 
be log-normally distributed if $X_0$ is log-normally 
distributed, or if $k$ or $\tau$ is {\it normally} 
distributed.  One possible configuration is to let the 
growth coefficient $k$ be the same for all plasma voids, 
with the duration of the growth spurt being normally 
distributed. Such a scenario would {\it a posteriori} 
justify the combination of SAD measurements from 16 
individual flares, since it presumes that all plasma voids 
grow at the same rate.  (The opposite arrangement --- 
normally distributed $k$ and uniform $\tau$ --- would also 
yield a log-normal $X(\tau)$; but it is more difficult to 
imagine all plasma voids having growth spurts of the same 
duration.)

In prior works \citep[and references 
therein]{SavageMcKenzie_11} SADs have been interpreted as 
being created in a burst of reconnection at a localized 
patch in the current sheet beneath the coronal mass 
ejection and above the flare arcade. In that vein it is 
presumed that the size and flux of a SAD is determined by 
the reconnection rate during that localized burst 
\citep[see also][]{LintonLongcope_06}. The scenario 
presented above then assigns $k$ to some parametrization 
of the reconnection rate, and leads to the supposition 
that $k$ might be uniform for all reconnecting patches. 
While this appears to run counter to the concept that the 
reconnection rate is determined by the local microphysics, 
we note that the simulations of \citet{Shay_EA07} yielded 
nearly uniform reconnection rates for a nontrivial range 
of system sizes, aspect ratios, and ion/electron mass 
ratios. Additional numerical simulations of patchy 
reconnection may determine whether the prescription 
suggested above yields distributions of SAD sizes and 
fluxes similar to those found in the observations.

\section{Conclusion}

Pursuant to the idea that the characteristics of 
reconnection products should yield clues about the 
reconnection process, we have compared the measured sizes 
and estimated magnetic fluxes of SADs to four theoretical 
distribution functions.  To avoid spurious variations in 
the distribution functions that may arise when data from 
multiple instruments are combined, we have examined only 
SADs measured with {\em Yohkoh}/SXT in 16 flares, at the 
expense of reduced spatial resolution and thereby reduced 
range in SAD sizes.  We find that the cross-sectional 
areas of 120 SADs appear to follow a log-normal 
distribution, while the estimated magnetic fluxes are 
consistent with either a log-normal or an exponential 
distribution. The data are incompatible with a power-law 
distribution, and thus do not appear to favor a fractal 
process for SAD creation.

The observed distribution of SAD sizes is likely affected 
by the processes of creation, fragmentation, and merging.  
Although fragmentation and merging may be possible for 
SADs, they have not been clearly observed in the coronal 
images to date. Future observations may provide evidence 
for fragmentation and/or merging, allowing empirical 
estimation of their relative importance.

In regards to creation of SADs, we offer a plausible 
scenario in which a log-normal distribution can be 
generated with minimal suppositions: Firstly, that the 
SADs form through patchy reconnection and grow at a rate 
described by equation (1).  Secondly, that the growth 
coefficient $k$ is approximately uniform for all SADs, 
while the duration of the growth is normally distributed.  
Although the conjecture of uniform $k$ appears to imply a 
uniform reconnection rate for all resistive patches in 
supra-arcade current sheets, this scenario remains to be 
tested via numerical simulations of reconnection.

\acknowledgments

This work was partially supported by NASA under contract 
NNM07AB07C with the Harvard-Smithsonian Astrophysical 
Observatory. {\em Yohkoh} data are provided courtesy of 
the NASA-supported {\em Yohkoh} Legacy Archive at Montana 
State University.  We gratefully acknowledge the helpful 
comments of an anonymous referee.


\clearpage

\begin{table}
\begin{center}
\caption{List of flares analyzed in this work.  \label{tbl-1}}
\begin{tabular}{lccr}
\tableline\tableline
Date & Time of   & GOES   & Number  \\
     & SADs (UT) & class. & of SADs \\
\tableline
23 Apr 1998  & 06:00-07:10 & X1.2 & 10 \\
27 Apr 1998  & 09:42-11:50 & X1.0 & 7 \\
09 May 1998  & 05:07-05:39 & M7.7 & 3 \\
30 Sep 1998  & 13:40-13:52 & M2.9 & 5 \\
23 Nov 1998  & 11:53-12:27 & M3.2 & 5 \\
20 Jan 1999  & 20:36-22:58 & M5.2 & 25 \\
03 May 1999  & 06:06-06:30 & M4.4 & 14 \\
25 Jul 1999  & 13:35-13:56 & M2.4 & 6 \\
28 Nov 1999  & 20:39-22:08 & M1.6 & 6 \\
22 Feb 2000  & 20:28-21:11 & M1.1 & 6 \\
12 Jul 2000  & 21:16-21:47 & M1.9 & 10 \\
08 Nov 2000  & 23:24-23:27 & M7.5 & 2 \\
03 Apr 2001  & 03:43-06:51 & X1.2 & 5 \\
26 Jun 2001  & 15:33-18:48 &  & 9 \\
30 Oct 2001  & 19:05-20:39 & C5.0 & 3 \\
01 Nov 2001  & 14:50-16:10 & M1.8 & 4 \\

\tableline
\end{tabular}
\end{center}
\end{table}

\clearpage

\begin{table}
\begin{center}
\caption{Goodness-of-fit parameters calculated via the Kuiper variant 
of the Kolmogorov-Smirnov test, for each of four distribution 
functions.  Small significance values (e.g., $\lesssim 0.10$) indicate 
that the CDF is not compatible with the data.  \label{tbl-2}}
\begin{tabular}{lcccc}
\tableline\tableline
Sample & Log-normal dist. & Power-law & Normal & Exponential \\
\tableline
SAD area  & D=0.097 & D=0.499 & D=0.187 & D=0.527 \\
 & sig=0.71 & sig=$3.9\times10^{-25}$ & sig=$5.7\times10^{-3}$ & sig=$3.2\times10^{-28}$ \\
\\
SAD flux  & D=0.128 & D=0.416 & D=0.333 & D=0.126 \\
 & sig=0.25 & sig=$3.8\times10^{-17}$ & sig=$1.4\times10^{-10}$ & sig=0.27 \\
\tableline
\end{tabular}
\end{center}
\end{table}

\end{document}